\documentclass[aps,prd,a4paper]{revtex4}
\usepackage{ulem}
\usepackage{color}
\usepackage{graphicx}
\usepackage{epsfig}
\usepackage{graphicx,epsfig,amsmath,bm,array,subfigure,amssymb}
\usepackage{slashed}
\usepackage{hyperref}

\begin{document}

\title{Heavy quark transport in a viscous semi QGP}
\author{Balbeer Singh$^{1,2}$, Hiranmaya Mishra$^1$}

\affiliation{$^{1}$Theory Division, Physical Research Laboratory, 
Navrangpura, Ahmedabad 380 009, India}

\affiliation{$^2$ Indian Institute of Technology Gandhinagar
	Gandhinagar 382 355, Gujarat, India}

\begin{abstract}
We study the effect of shear and bulk viscosities on the heavy quark transport coefficient within the matrix model of semi QGP. Dissipative effects are incorporated through the first-order viscous correction in the quark/antiquark and gluon distribution function. It is observed that while the shear viscosity effects reduces the drag of heavy quark the bulk viscosity effects increase  the drag and the diffusion coefficients of heavy quark. For finite values of $\eta/s$ and $\xi/s$, Polyakov loop further decreases the drag and the diffusion coefficients as compared to perturbative QCD. 
\end{abstract}

\maketitle
\section{Introduction}
The aim of heavy ion collision (HIC) experiments is to characterize the properties of the deconfined state of matter namely the  quark-gluon plasma (QGP) which is being created in these collisions. In this regard, the energy loss of heavy quark (HQ); especially charm and bottom; in the QGP medium is considered as one of the promising probes of QGP. There are two mechanisms that contribute to the energy loss of HQ; one is the radiative energy loss (medium induced gluon radiation) and the other is the collisional energy loss (i.e., scattering of HQ with thermalized medium partons). At low energy, the dominant contribution to the energy loss comes from the collision processes. The in-medium energy loss of HQ is manifested  in the large elliptic flow i.e., $v_2$ and in the suppression of high momentum heavy flavored (HF) hadrons as compared to proton-proton collision~\cite{Adler:2003kt, Adler:2002ct, Adare:2006nq, Abelev:2006db, Aaboud:2018bdg,Dong:2018fhv,Rapp:2018qla}.

Heavy quarks are produced in the initial stages of the collisions during the hard scatterings governed by perturbative quantum chromodynamics (pQCD) mostly through gluon fusion~\cite{Satz:2018oiz}; for next-to-leading order production see Ref.~\cite{Cacciari:1998it,Cacciari:2012ny}. Because of the large mass of HQ as compared to the temperature  ranges accessible in the Relativistic Heavy Ion Collider (RHIC) and the Large Hadron Collider (LHC) energies, the thermal production of HQ is negligible. Hence, once produced in the hard collisions, HQ propagates throughout the space-time evolution of the medium and interact with the light thermal partons (light quarks and gluons). Thus, the resulting effect of the interaction of HQ with the bulk medium modifies the spectra of HF hadrons. The interaction of HQ with the bulk medium is described by the scattering of HQ with the light thermal partons of the medium. At low momentum, the dominant contribution to the HQ scattering off of light quark and gluon in the thermal medium comes from the elastic scatterings and can be described by the diffusion process akin to Brownian motion. In addition, the thermalization of HQ in the bulk medium is also slowed down due to its large mass. Hence, the transport of non-equilibrated HQ in the thermalized medium of light quark and gluons yield valuable information about the medium throughout its propagation. In particular, while  the low momentum interaction of HQ with the bulk medium is characterized by the spatial diffusion coefficient, the energy loss of HQ is described by the drag coefficient.	

The electron-positron yield associated with the semileptonic decays of HF meson with $1.2<p_{T}<10$ GeV in Au-Au collision at RHIC shows a strong suppression,  indicating a significant energy loss of HQ in the medium~\cite{Abelev:2006db, Adare:2006nq}.  Furthermore, a large collective flow of heavy quark for $0.5<p_{T}<5$ GeV has also been reported in Ref.~\cite{Adare:2006nq}. Perturbative QCD based calculations for HQ energy loss cannot explain the observed suppression and collective flow ~\cite{Wicks:2005gt} so it is required to include possible non-perturbative effects. There have been various efforts to incorporate non-perturbative effects using various models such as T-matrix model\cite{Mannarelli:2005pz, Liu:2017qah}, quasi particle model~\cite{Song:2015sfa,Berrehrah:2013mua, Das:2015ana}, resonance model ~\cite{vanHees:2004gq} for estimating the HQ transport. In Ref.~\cite{vanHees:2007me}, HQ transport coefficients are evaluated in T-matrix approach including non-perturbative effects by employing the potential interaction of heavy-light quark extracted from lattice QCD simulations. A good agreement with the observed $R_{AA}$ and collective flow $v_2$ of this calculation suggests existance of the strongly interacting nature of QGP fluid produced in RHIC.  In fact, from lattice simulations, it has been argued that the non-perturbative effects are significant a temperature up to twice of pseudo critical temperature i.e., $T_c\sim 160$ MeV ~\cite{Petreczky:2004pz, Bazavov:2011nk}.  Recently, based on a Polyakov loop model calculation, heavy quark drag and diffusion coefficients have been computed for charm quark in Ref.~\cite{Singh:2018wps}. Similar to Ref.{\cite{vanHees:2007me,Mannarelli:2005pz}}, the transport coefficients temperature and momentum dependence have appeared different as compared to perturbative calculations. The drag coefficient are observed being rather flat regarding temperature dependence while diffusion coefficient exibit a strong temperature dependenc.  The consistency in the results suggest that there could be some model independent correlations between the results obtained within the Polyakov loop and other non-perturbative models from a different perspectives.

On the other hand, QGP, formed in HICs, behaves  like almost an ideal fluid with a very small value for the ratio of shear viscosity to entropy density $\eta/s$. Evidence for such a small $\eta/s$ is provided by the large elliptic flow data that requires $\eta/s\sim 0.08-0.2$~\cite{Qiu:2011hf,Romatschke:2007mq,Bhalerao:2015iya,Schenke:2011bn}. Viscous coefficients in the QGP as well as in the hadronic medium have been studied in Refs.\cite{Prakash:1993bt,Baym:1990uj,Arnold:2003zc}. In these studies it was found that the dominant contribution of dissipation in both the QGP and the hadronic medium arises from shear viscosity. However, bulk viscosity is equally important and may dominates near transition temperature i.e., $\xi/s\sim 1$ ~\cite{Kharzeev:2007wb} and can singnificantly affect hadron $p_{T}$  spectra and elliptic flow $v_2$~\cite{Dusling:2011fd}. Viscous corrections have also been studied for dilepton production in QGP~\cite{Vujanovic:2013jpa,Bhatt:2011kx}, photon production~\cite{Bhatt:2010cy}, damping rate of heavy quark~\cite{Sarkar:2012fk}, heavy quark radiative energy loss~\cite{Sarkar:2018erq,Baier:2006pt,Baier:2000mf}, event-plane correlations~\cite{Chattopadhyay:2017bjs,Qiu:2012uy} etc. Effect of shear and bulk viscosity on HQ drag and diffusion coefficient have been studied in Ref.\cite{Das:2012ck} using a fugacity model. In the present study, we intend to include the viscous corrections (both shear and bulk) along with a non-trivial Polyakov loop background that is used to describe the ``semi QGP" within a matrix model.  We find that in the perturbative limit our results are consistant with the previous results, however, with the inclusion of Polyakov loop ($\phi$), at low temperature our results are different from that of Ref.\cite{Das:2012ck}.  In this work, we include the viscous corrections (both shear and bulk) in the single particle distribution functions of quark and gluon to estimate the viscous effects on the HQ transport coefficients. We estimate this using Fokker-Plank equation and use the matrix model of semi QGP to evaluate the relavant scattering amplitudes. The single particle distribution function (see Eqs.(\ref{quarkdis}) and (\ref{gluondis})) is modified using second moment ansatz. In Ref.~\cite{Sarkar:2018erq} it was shown that viscous effects induce a larger energy loss of HQ. So one may expect that, viscous corrections may be important and significantly affect the transport properties of HQ in the bulk medium. However, we find that for small shear and bulk viscosities, the dissipative effects on the drag and the diffusion coefficients are somewhat weak.  

We organize this work as follows. An introduction on the formalism for evaluation of HQ drag and diffusion coefficients within the matrix model of semi QGP is discussed in section (\ref{formalism}) which is followed by the discussion on semi QGP in section (\ref{semiqgp}). In this section, we also discuss some salient features of the matrix model. In section (\ref{viscousdist}), an ansatz for the first order viscous correction on quark/gluon distribution is discussed. In section (\ref{scattering}), we discuss the interaction of HQ with the light thermal parton and present matrix element squared for Coulomb and Compton scatterings within the matrix model. These matrix element squared are used to evaluate the drag and diffusion coefficients . Finally, in section (\ref{results}) we discuss the  viscous effects on HQ quark transport and present the numerical results for the drag and the diffusion coefficient for constant values of $\eta/s$ and $\xi/s$. Finally, we summarise and give an outlook of the present work in section(\ref{summary}).  

\section{Formalism}
\label{formalism}
The Brownian motion of HF particles can be described by the Fokker-Plank equation where the interactions of heavy quark with the bulk of light quarks and gluons are encoded in the transport coefficient. Assuming that HF quark of momentum $\boldsymbol{p}$ is traveling in a medium of light quark and gluon, the Boltzamann equation for phase-space distribution $f_{Q}$ of heavy quark can be written as \cite{Svetitsky:1987gq}
\begin{equation}
\bigg[\frac{\partial}{\partial t}+\frac{\boldsymbol{p}}{E_p}\frac{\partial}{\partial \boldsymbol{x}}+\boldsymbol{F}\frac{\partial}{\partial \boldsymbol{p}}\bigg]f_Q{(\boldsymbol{p},\boldsymbol{x},t)}= C[f_{Q}],
\label{boltzmann}
\end{equation} 
where $\boldsymbol{F}$ is the force due to external mean-field such as chromo electric or magnetic fields present in the intial stages of the heavy ion-collision , $E_p=\sqrt{m_Q^2+\boldsymbol{p}^2}$ is the energy of heavy quark with mass $m_Q$ and $C[f_Q]$ is the collision integral. Neglecting the mean-field effects, Eq[\ref{boltzmann}] reduces to
\begin{equation}
\frac{\partial}{\partial t}f_Q{(\boldsymbol{p},t)}= C[f_{Q}].
\label{boltzmann1}
\end{equation}
On the right-hand side of Eq.[\ref{boltzmann1}], collision integral in  terms of collision rate which change the momentum of HF quark from $\boldsymbol{p}$ to $\boldsymbol{p}-\boldsymbol{k}$  is written as
\begin{equation}
C[f_{Q}]=\int d^3k[w(\boldsymbol{p}+\boldsymbol{k},\boldsymbol{k})f_{Q}(\boldsymbol{p}+\boldsymbol{k})-w(\boldsymbol{p},\boldsymbol{k})f_{Q}(\boldsymbol{p})],
\label{trate}
\end{equation}
where $w$ is the transition rate of heavy quark colliding with heat bath particles of momentum $\boldsymbol{k}$. The first term in Eq.[\ref{trate}] is the gain term that describes the transition of HF quark from a state of momentum $\boldsymbol{p}+\boldsymbol{k}$ to momentum state $\boldsymbol{p}$ while the loss term (second term) represents the scattering out from the momentum state $\boldsymbol{p}$. Assuming the scatterings of HF quark with the bulk medium partons is dominated by small momentum transfer i.e., $|\boldsymbol{k}| \ll |\boldsymbol{p}|$, the distribution function of HQ and transition rate can be expanded up to second order with respect to $\boldsymbol{k}$ .ie.,
\begin{equation}
w(\boldsymbol{p}+\boldsymbol{k},\boldsymbol{k})f_{Q}(\boldsymbol{p}+\boldsymbol{k}) \simeq w(\boldsymbol{p},\boldsymbol{k})f_{Q}(\boldsymbol{p})+\boldsymbol{k}\frac{\partial}{\partial \boldsymbol{p}}[w(\boldsymbol{p},\boldsymbol{k})f_{Q}(\boldsymbol{p})]+\frac{1}{2}k_{i}k_{j}\frac{\partial^2}{\partial p_{i}\partial p_{j}}[w(\boldsymbol{p},\boldsymbol{k})f_{Q}(\boldsymbol{p})].
\end{equation} 
With this approximation the collision integral simplifies to 
\begin{equation}
C[f_Q]=\int d^3k \bigg[k_{j}\frac{\partial}{\partial p_{j}}+\frac{1}{2}k_{i}k_{j}\frac{\partial^2}{\partial p_{i}\partial p_{j}}\bigg]w(\boldsymbol{p},\boldsymbol{k})f_{Q}(\boldsymbol{p}).
\end{equation}
The function $w$ can be expressed in terms of the cross-section for scattering processes in the heat bath. For, scattering of HQ with momentum $\boldsymbol{p}$ with the bulk medium-light thermal parton with momentum $\boldsymbol{q}$, one finds
\begin{equation}
w(\boldsymbol{p},\boldsymbol{k})=\gamma_{l}\int \frac{d^3q}{(2\pi)^3} f(q)_{l} |\boldsymbol{v}_{rel}| \frac{d\sigma}{d\Omega}(\boldsymbol{p},\boldsymbol{q}\rightarrow \boldsymbol{p}-\boldsymbol{k},\boldsymbol{q}+\boldsymbol{k}),
\end{equation}
where $f(q)_{l}$ is Fermi-Dirac/or Bose-Einstein distribution function of light thermal partons and $\gamma_{l}$ is degeneracy factor which is $\gamma_{q}=6$ for quarks and $\gamma_{g}=16$ for gluons. Boltzmann equation Eq.[\ref{boltzmann1}] can be approximated as Fokker-Plank equation
\begin{equation}
\frac{\partial }{\partial}f_{Q}(\boldsymbol{p},t)=\frac{\partial}{\partial p_{i}}\bigg(A_{i}(\boldsymbol{p})f_{Q}(\boldsymbol{p},t)+\frac{\partial}{\partial p_{j}}B_{ij}(\boldsymbol{p})f_{Q}(\boldsymbol{p},t)\bigg).
\end{equation}
Here $A_{i}$ and $B_{ij}$ are drag and diffusion coefficient and are given as
\begin{equation}
A_{i}(\boldsymbol{p})=\int d^3k w(\boldsymbol{p},\boldsymbol{k}) k_{i}
\end{equation} 
\begin{equation}
B_{ij}(\boldsymbol{p})=\frac{1}{2}\int d^3k w(\boldsymbol{p},\boldsymbol{k}) k_{i}k_{j}.
\end{equation}
For an isotropic heat bath at local thermal equiliberium one may define \cite{Rapp:2009my} 
\begin{equation}
A_{i}(\boldsymbol{p})=A(\boldsymbol{p})p_{i},
\end{equation}
\begin{equation}
B_{ij}(\boldsymbol{p})=B_{0}(\boldsymbol{p})P^{\parallel}_{ij}+B_{1}(\boldsymbol{p})P^{\perp}_{ij},
\end{equation}
where $P^{\parallel}_{ij}$ and $P^{\perp}_{ij}$ are longitudinal and transver projection operators defined as
\begin{equation}
P^{\parallel}_{ij}=\frac{p_{i}p_{j}}{|\boldsymbol{p}|^2}\hspace{0.1cm}       ,\hspace{1cm}P^{\perp}_{ij}=\delta_{ij}-\frac{p_{i}p_{j}}{|\boldsymbol{p}|^2}.
\end{equation}
For a process $l Q \rightarrow l Q$ (where $l$ stands for light quarks and gluon) the drag and diffusion coefficients of HQ in the plasma of light quarks and gluons are given by the scalar integral of form
\begin{eqnarray}
\langle X(\boldsymbol{p'}) \rangle&=&\frac{1}{2 E_p}\int \frac{d^3q}{(2\pi)^3 2E_q} \int \frac{d^3p'}{(2\pi)^3 2E_{p'}}\int \frac{d^3q'}{(2\pi)^3 2E_{q'}}|\mathcal{M}|^2 \nonumber\\
&\times&(2\pi)^4 \delta^{4}(p+q-p'-q')f_{l}(q)(1\pm f_{l}(q))X(\boldsymbol{p'}),
\label{integral}
\end{eqnarray}
where $l=q,\bar{q},g$. In the present study, we evaluate scattering amplitude for relavant $2\rightarrow 2$ processes within the matrix model which make the matrix element squared color dependent.  So in the presence of a background gauge field Eq.(\ref{integral}) becomes
\begin{eqnarray}
\langle X(\boldsymbol{p'}) \rangle&=&\frac{1}{2 E_p}\int \frac{d^3q}{(2\pi)^3 2E_q} \int \frac{d^3p'}{(2\pi)^3 2E_{p'}}\int \frac{d^3q'}{(2\pi)^3 2E_{q'}}\bigg(\sum_{a,e}|\mathcal{M}_{qQ}|_{ab}^2 f_{a}(q)(1-f_{e}(q'))\nonumber \\
&+&\sum_{e,f,g,h}|\mathcal{M}_{gQ}|_{efgh}^2f_{ef}(q)(1+f_{gh}(q'))\bigg) (2\pi)^4 \delta^{4}(p+q-p'-q')X(\boldsymbol{p'}),
\label{integral1}
\end{eqnarray}
where $a,e$ are color indices of incoming and outgoing light quark and $ef,gh$ are color indices for incoming and outgoing gluon that interact with HQ, $|\mathcal{M}_{qQ}|_{ab}^2$ and $|\mathcal{M}_{gQ}|_{efgh}^2$ are matrix element squared respectively for the processes $q^{a}Q^{c}\rightarrow q^{b}Q^{d}$ and $g^{ef}Q^{a}\rightarrow g^{gh}Q^{b}$. In the notation as written in Eq.[\ref{integral1}], the drag and diffusion coefficients are written as
\begin{equation}
A(\boldsymbol{p})=\langle 1 \rangle - \frac{\langle\boldsymbol{p}\cdot \boldsymbol{p'}\rangle}{|\boldsymbol{p}|^2} 
\label{drag}
\end{equation}
\begin{equation}
B_{0}(\boldsymbol{p})=\frac{1}{4}\bigg(\langle |\boldsymbol{p'}|^2 \rangle-\frac{\langle(\boldsymbol{p}\cdot \boldsymbol{p'})^2\rangle}{|\boldsymbol{p}|^2} \bigg) 
\label{diffusion}
\end{equation}
\begin{equation}
B_{1}(\boldsymbol{p})=\frac{1}{2}\bigg(\frac{\langle(\boldsymbol{p}\cdot\boldsymbol{p'})^2\rangle}{|\boldsymbol{p}|^2}-2\langle \boldsymbol{p}\cdot \boldsymbol{p'}\rangle+|\boldsymbol{p}|^2\langle 1\rangle\bigg).
\end{equation}
In the presence of a non-trivial Polyakov loop background, apart from the matrix elements, the distribution functions also become color dependent. We evaluate these scattering amplitudes in the matrix model of semi QGP  which we discuss in the next section.  
\section{Semi QGP}
\label{semiqgp}
At high temperature, the density of colored particles like quarks and gluon are large and can be calculated using perturbative QCD. However, at low temperature, colored particles are statistically suppressed and are measured by the small value of Polyakov loop e.g., at chiral cross-over temperature $T_c\sim 170$ MeV, $\phi=0.2$~\cite{Bazavov:2016uvm} which is way smaller from its asymptotic value i.e., $\phi=1$. Because of the suppression of colored particles, the region near chiral cross-over is termed as semi-QGP \cite{Pisarski:2016ixt}. Semi QGP is characterized by the Polyakov loop as defined in Eq.(\ref{poly1}). For the calculational purpose, we shall use double line notation which is quite useful in the matrix model of semi QGP. In the double line basis, quark carries one color index say $a=1,2,..,N$ and gluons carry double index say $ab=1,2,..,N^2$. For $SU(N)$ group such $N^2$ pairs lead to $N^2$ generators and the basis is overcomplete by one generator. The overcomplete basis is compensated by introducing the projection operator defined as \cite{Hidaka:2009hs,tHooft:1973alw,Cvitanovic:1976am} 
\begin{equation}
\mathcal{P}^{ab}_{cd}=\mathcal{P}_{ba;cd}=\mathcal{P}^{ab;dc}=\delta^{a}_{c}\delta^{b}_{d}-\frac{1}{N}\delta^{ab}\delta_{cd}
\label{projection}
\end{equation}
hence the generator is given by
\begin{equation}
(t^{ab})_{cd}=\frac{1}{\sqrt{2}}\mathcal{P}^{ab}_{cd}.
\label{pop}
\end{equation}
The trace over two generatros doesn't vanish but rather is again a projection operator i.e.,
\begin{equation}
Tr(t^{ab} t^{cd})=\frac{1}{2}\mathcal{P}^{abcd}.
\end{equation}
This is due to the presence of extra generator as compared to generators in an orthonormal basis. The structure constant of the group in the double line basis is given by
\begin{equation}
f^{ab,cd,ef}=\frac{i}{\sqrt{2}}(\delta^{ad}\delta^{cf}\delta^{eb}-\delta^{af}\delta^{cb}\delta^{ed}).
\label{structure}
\end{equation}
In the mean-field approximation, the constant background field is defined as $A_{\mu}^{0}=\frac{1}{g}\delta_{\mu 0}Q^{a}$ with $Q^{a}=2 \pi q^{a} T$. Since $A_{0}$ is traceless so sum over $Q$'s
vanishes i.e., $\sum_{a} Q^{a}=0$. For an $SU(3)$ group, $Q^{a}=(-Q^{i},-Q^{i-1},..0,Q^{i-1},Q^{i})$, where $i=N/2$ if $N$ is even and $(N-1)/2$ if $N$ is odd. 
In the temporal direction, the Wilson line is written as
\begin{equation}
P=\mathcal{P} \exp\bigg(ig\int_{0}^{\beta}d\tau A_{0}(x_{0},\boldsymbol{x})\bigg)
\end{equation}
where $\mathcal{P}$ stands for the ordering of imaginary time and $\tau$ is imaginary time. Polyakov loop, which is the trace of Wilson line, in the constant background gauge field can be written as
\begin{equation}
\phi=\frac{1}{N}\sum_{a=1}^{N} \exp(i 2\pi q^{a}).
\label{poly}
\end{equation} 
For an $SU(3)$ group, where $q^{a}=(-q,0,q)$ Eq.[\ref{poly}] is simplified to
\begin{equation}
\phi=\frac{1}{3}(1+2\cos(2 \pi q)).
\label{poly1}
\end{equation}
The background gauge field acts as an imaginary chemical potential for colored particles so the statistical distribution function of quark/anti-quark and the gluon are 
\begin{equation}
f^{0}_{a}(E)=\frac{1}{e^{\beta(E-iQ_{a})}+1}, \hspace{1cm} \tilde{f}^{0}_{a}(E)=\frac{1}{e^{\beta(E+iQ_{a})}+1},
\label{qrkdist}
\end{equation}
\begin{equation}
f^{0}_{ab}(E)=\frac{1}{e^{\beta(E-i(Q_{a}-Q_{b}))}-1},
\label{gludist}
\end{equation}
where the single and double indices are for quark/antiquark and gluon. For a background field and given $Q^a$ these distribution functions are complex so are unphysical. Physical meaning comes when one integrates over all distributions of $Q^{a}$.  The resummed gluon propagator in the presence of a static background gauge field is given as ~\cite{Hidaka:2009ma}   
\begin{equation}
D{\mu \nu ; a b c d}(K)=P^{L}_{\mu \nu} \frac{k^2}{K^2} D^{L}_{a b c d}(K)+P^{T}_{\mu \nu}D^{T}_{a b c d}(K),
\label{propef}
\end{equation}
where $P^T_{\mu \nu}=g_{\mu i}\bigg(-g^{i j}-\frac{k^{i} k^{j}}{K^2}\bigg)g_{j\nu}$ and $P^L_{\mu \nu}=-g_{\mu \nu}+\frac{k_{\mu}k_{\nu}}{K^2}-P^{T}_{\mu \nu}$ are the longitudinal and the transverse projection operators. The longitudinal and the transverse gluon propagators are written as
\begin{equation}
D^{L}_{\mu \nu ; a b c d}(K)=\bigg(\frac{i}{K^2+F}\bigg)_{a b c d},
\end{equation}
\begin{equation}
D^{T}_{\mu \nu ; a b c d}(K)=\bigg(\frac{i}{K^2-G}\bigg)_{a b c d},
\end{equation}
where 
\begin{equation}
F=2 M^2\bigg(1-\frac{y}{2}\ln\bigg(\frac{y+1}{y-1}\bigg)\bigg),
\end{equation}
\begin{equation}
G= M^2\bigg(y^2+\frac{y(1-y^2)}{2}\ln\bigg(\frac{y+1}{y-1}\bigg)\bigg),
\end{equation}
with $y=\frac{k_{0}}{|\boldsymbol{k}|}$ and $M^2=(M^2)_{a b c d}$ is the thermal mass of the gluon. For the drag and the diffusion of HQ studied here, the momentum transfer is small so only longitudinal propagator contributes to the squared matrix elements~\cite{Rapp:2009my,Moore:2004tg} 

\section{Viscous corrections in the distribution functions}
\label{viscousdist}
In this section, we briefly describe the first order viscous corrections on the thermal distribution function of quarks and gluons. We start with the energy-momentum tensor of a non-ideal fluid which is given as~\cite{Dusling:2011fd}
\begin{equation}
T^{\mu \nu}= (\epsilon+P)u^{\mu}u^{\nu}+P g^{\mu \nu}+\pi^{\mu \nu}+\Pi \nabla^{\mu \nu},
\label{emten}
\end{equation}
where $\epsilon, P, u^{\mu}$ are the energy density, pressure density and four-velocity of the fluid. For metric tensor, we use the convention $g^{\mu \nu}=diag(-1,+1,+1,+1)$ so that $u^{\mu}u_{\mu}=-1$ and the term  $\nabla^{\mu \nu}=g^{\mu \nu}+u^{\mu}u^{\nu}$. The first two terms at the right hand side of Eq.(\ref{emten}) describes the energy-momentum tensor for an ideal fluid  and the rest two terms are part of viscous corrections that summarises the effect of shear and bulk viscosities respectively. The dissipative terms are constructed from the derivatives $\Delta^{\alpha}=\nabla^{\alpha \beta}\partial_{\beta}$ and $\nabla^{\mu \nu}$.  In the first-order approximation, the symmetric tensor $\pi^{\mu \nu}$ satisfying the condition $u_{\mu}\pi^{\mu \nu}=0$, in the local rest frame is given as
\begin{equation}
\pi^{\mu \nu}=-\eta \bigg(\Delta^{\mu}u^{\nu}+\Delta^{\nu}u^{\mu}-\frac{2}{3}\nabla^{\mu \nu}\Delta_{\alpha}u^{\alpha}\bigg)
\label{shear}
\end{equation}
and the bulk viscosity dependent term
\begin{equation}
\Pi=-\xi \Delta_{\alpha}u^{\alpha}.  
\label{bulk}
\end{equation}
Dissipative effects can be incorporated in the color dependent distribution functions $f_{a/ab}(E)$ which contains the ideal part as well as viscous corrections. For this purpose, we write   $f_{a/ab}(E)=f^{0}_{a/ab}(E)+\delta f_{a/ab}(E)$ ($f^{0}_{a/ab}(E)$ is equilibrium distribution function of quark/antiquark and gluon) and use the second-moment ansatz as in Refs.\cite{Teaney:2003kp,Dusling:2007gi,Dusling:2011fd}, so that
\begin{equation}
\delta f(E)_{a/ab}=\frac{1}{T^3 s} f(E)^{0}_{ a/ab} (1+f(E)^{0}_{ a/ab})p^{\mu}p^{\nu} \bigg(\frac{A}{2}\pi_{\mu \nu}+\frac{B}{5}\Pi \nabla_{\mu \nu}\bigg) 
\end{equation}
where $A$ and $B$ are constants. Constrain on $\delta f_{a/ab}$ comes from the continuity of stress-energy tensor across the freeze-out hypersurface~\cite{Dusling:2007gi} i.e.,
\begin{equation}
\delta T^{\mu \nu}=\int \frac{d^3 k}{(2\pi)^3}\frac{k^{\mu}k^{\nu}}{E_k}\delta f_{a/ab}(E).
\end{equation}
The choice of $\delta f_{a/ab}$ is not unique, as pointed out in Ref.\cite{Dusling:2009df}, $\delta f_{a/ab}$ can have linearly increasing form with momentum and also quadratically increasing with momentum or anything in between linear to quadratic increasing behavior. However, we will continue with the form as in Ref.\cite{Dusling:2007gi,Dusling:2011fd}.  In the local rest frame of the fluid where $u_{0}=1, u_{i}=0, \partial_{\mu}u_{0}=0$ and $\partial_{\mu}u_{i}\neq 0$, the deviation in distribution function can be written as \cite{Dusling:2007gi,Dusling:2011fd}
\begin{equation}
\delta f(E)_{a/ab}=\frac{1}{T^3 s} f(E)^{0}_{ a/ab} (1\mp f(E)^{0}_{ a/ab})p^{\mu}p^{\nu} \bigg(\frac{1}{2}\pi_{\mu \nu}+\frac{1}{5}\Pi \nabla_{\mu \nu}\bigg). 
\label{distfn}
\end{equation}
With further simplification using Eqs.[\ref{shear}] and [\ref{bulk}], distribution functions of quark and gluon becomes
\begin{equation}
f(E)_{a}=f(E)^{0}_{a}+\frac{f(E)^{0}_{a}(1-f(E)^{0}_{a})}{T^3 \tau} \bigg[\frac{\eta}{s}\bigg(-p_{z}^2+\frac{{p}^2}{3}\bigg)+\frac{\xi}{s}\frac{p^2}{5}\bigg]
\label{quarkdis}
\end{equation}
\begin{equation}
f(E)_{ab}=f(E)^{0}_{ab}+\frac{f(E)^{0}_{ab}(1+f(E)^{0}_{ab})}{T^3 \tau} \bigg[\frac{\eta}{s}\bigg(-p_{z}^2+\frac{{p}^2}{3}\bigg)+\frac{\xi}{s}\frac{p^2}{5}\bigg]
\label{gluondis}
\end{equation}
where $\tau$ is the thermalization time. In the present investigation, for evaluating drag and diffusion coefficients, we shall use Eq.[\ref{quarkdis}] and Eq.[\ref{gluondis}] for quark/antiquark and gluon distribution function.
\section{Scatterings amplitudes within matrix model}
\label{scattering}
In this section, We shall discuss the scattering of HQ of mass $M$ and energy $E=\sqrt{p^2+M^2}$ with the light thermal partons in the bulk medium and we shall also compute the scattering amplitude squared within the matrix model of semi QGP. To compute the drag and the diffusion coefficients of HQ transport we shall follow a similar approach to include  screening effects as in Ref.\cite{Svetitsky:1987gq,Moore:2004tg}.  For the elastic collision, there are two types of scattering processes that contributes to the drag and the diffusion coefficient of HQ. One is Coulomb scattering i.e., scattering off of HQ with light quark and another is Compton scattering i.e., scattering off of HQ with gluons. In the following we present these in detail.

\begin{figure}[tbh]
\subfigure{
\includegraphics[width=3.5cm]{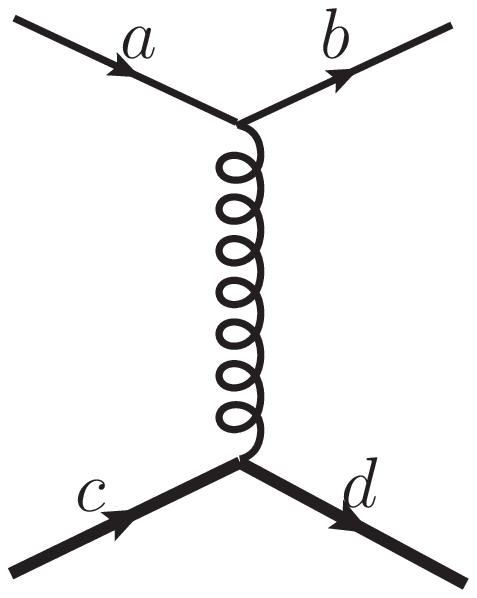}}
\subfigure{
\hspace{-0mm}\includegraphics[width=3.5cm]{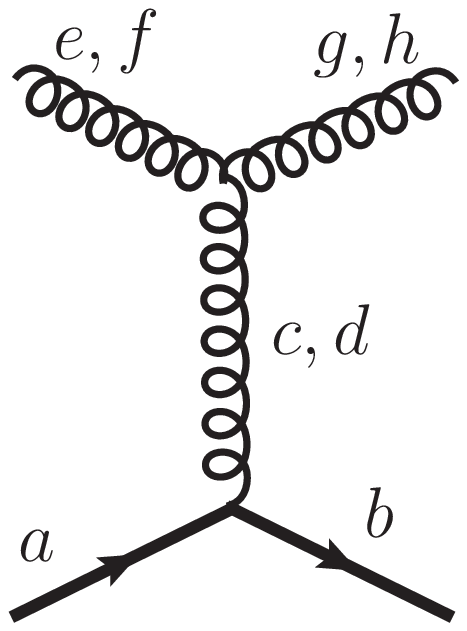}} 
\caption{Coulomb scattering (left) of HQ (bold solid line) and light quark/antiquark (thin solid line). t-channel Compton scattering (right). The curly line represent a gluon. }
\label{st}
\end{figure}

\textbf{Coulomb scattering:} The Feynmann diagram for the Coulomb scattering of HQ and a light quark is shown on the left side of Fig.(\ref{st}). Here $a, c, b, d$ are the color indices of initial and final quarks.  In the double line notation, the scattering amplitude for this process is 
\begin{equation}
i\mathcal{M}_{qQ}=\frac{(ig)^2}{(t+(m_D^2)_{mljk})}(t^{jk})_{ab}(t^{ml})_{cd}[\bar{u}_{b}(q')\gamma^{\mu}u_{a}(q)][\bar{u}_{d}(p')\gamma_{\mu}u_{c}(p)],
\label{coulomb}
\end{equation}
where $g$ is strong coupling constant, $t$ is Mandelstam variable and $m, l, j, k$ are the color indices of gluon propagator. In the limit of soft momentum transfer, only time-like component of the propagator contributes and the propagators simply become Debye screened propagator with $1/t \rightarrow 1/(t+m_D^2)$~\cite{Svetitsky:1987gq,Moore:2004tg}  where $m_D^2$ is color dependent Debye mass and can be given as
\begin{eqnarray}
(m_D^2)_{abcd}&=&\frac{g^2}{6}\bigg[\delta_{ad}\delta_{bc}\bigg(\sum_{e=1}^{3}\bigg(\mathcal{D}(Q_{ae})+\mathcal{D}(Q_{eb})\bigg)-N_f(\mathcal{\tilde{D}}(Q_{a})+\mathcal{\tilde{D}}(Q_{b}))\bigg)\nonumber\\
&-&2\delta_{ab}\delta_{cd}\bigg(\mathcal{D}(Q_{ac})-\frac{N_f}{N}\bigg(\mathcal{\tilde{D}}(Q_{a})+\mathcal{\tilde{D}}(Q_{c})\bigg)+\frac{N_f}{N^2}\sum_{e=1}^{3}\mathcal{\tilde{D}}(Q_{e})\bigg)\bigg],
\end{eqnarray}
where
\begin{equation}
\mathcal{D}(Q_{a})=\frac{3}{\pi^2}\int_{0}^{\infty} dE E\bigg(\frac{1}{e^{\beta(E+iQ_{a})}-1}+\frac{1}{e^{\beta(E-iQ_{a})}-1}\bigg),
\end{equation}
and $\mathcal{\tilde{D}}(Q_{a})=\mathcal{D}(Q_{a}+\pi T)$. In the perturbative limit, Eq.(\ref{coulomb}) can be written as
\begin{equation}
i\mathcal{M}_{qQ}=-\frac{g^2}{t}(t^{jk})_{ab}(t^{jk})_{cd}[\bar{u}_{b}(q')\gamma^{\mu}u_{a}(q)][\bar{u}_{d}(p')\gamma_{\mu}u_{c}(p)],
\label{coulomb1}
\end{equation}
and the product of projection operator with open color index $a, b$ can be written as
\begin{equation}
\mathcal{P}^{jk}_{ab}\mathcal{P}^{jk}_{cd}\mathcal{P}^{j'k'}_{ba}\mathcal{P}^{j'k'}_{dc}=(N-1)\bigg(1-\frac{\delta_{ba}}{N}\bigg).
\label{projection1}
\end{equation}
However, for the computation of the drag and the diffusion coefficient, we shall use Eq.(\ref{coulomb}). Simplifying Eq.(\ref{coulomb}) for massless light quark and massive heavy quark by summing and averaging over final and initial spins, the scattering amplitude squared ($|\mathcal{M}_{qQ}|^2$) can be written as
\begin{equation}
|\mathcal{M}_{qQ}|^2=\frac{g^4}{16N_c^2}\mathcal{P}^{jk}_{ab}\mathcal{P}^{ml}_{cd}\mathcal{P}^{j'k'}_{ba}\mathcal{P}^{m'l'}_{dc}
\frac{(8(s-M^2)^2+8(u-M^2)^2+16M^2t)}{(t+(m_D^2)_{mljk})(t+(m_D^2)_{m'l'j'k'})}
\label{coulomb3}
\end{equation}
Let us note here that the drag and the diffusion coefficient of HQ as defined in Eqs.(\ref{drag}) and (\ref{diffusion}) depends on the color of incoming and outgoing light quark i.e.,  $Q^{a}$ and $Q^{b}$ in the distribution functions; see Eq.(\ref{integral1}). So to compute the color-averaged quantity,  the color index $a$ and $b$ in Eq.(\ref{coulomb3}) will be summed with the distribution function. With the distribution function as defined in Eq.(\ref{quarkdis}), for Coulomb scattering the bracketed quantity in Eq.(\ref{integral1})   gives
\begin{eqnarray}
\langle X(p')\rangle&=&\frac{1}{2 E_{p}}\sum_{abcd}\int\frac{d^3q}{2E_{q}(2\pi)^3}\frac{d^3p'}{2E_{p'}(2\pi)^3}\frac{d^3q'}{2E_{q'}(2\pi)^3}\sum_{jkj'k'}\sum_{mlm'l'}|\mathcal{M}_{qQ}|^2(f^{0}_{a}(q)+\delta f_{a}(q))\nonumber\\
&\times&(1-f^{0}_{b}(q')-\delta f_{b}(q'))\langle X(p')\rangle 
\end{eqnarray}	
 
\begin{figure}[tbh]
\subfigure{
\includegraphics[width=4.5cm]{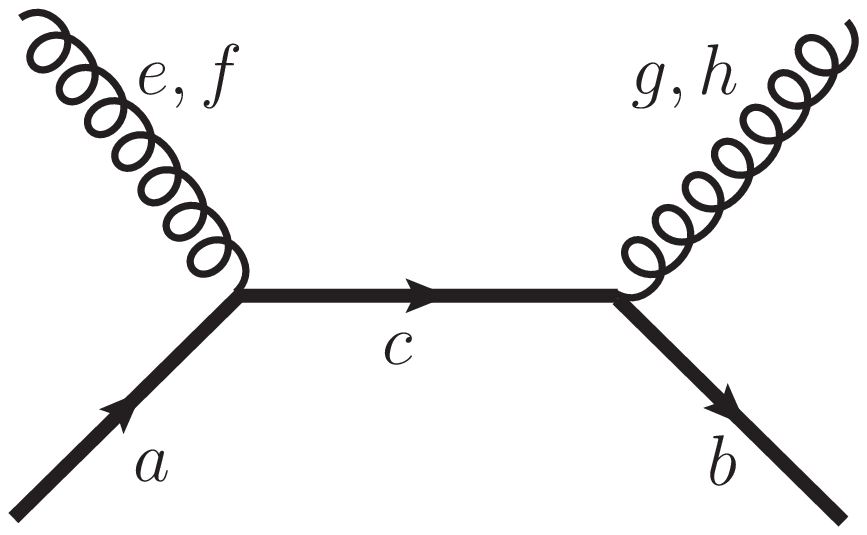}}
\subfigure{
\hspace{-0mm}\includegraphics[width=4.5cm]{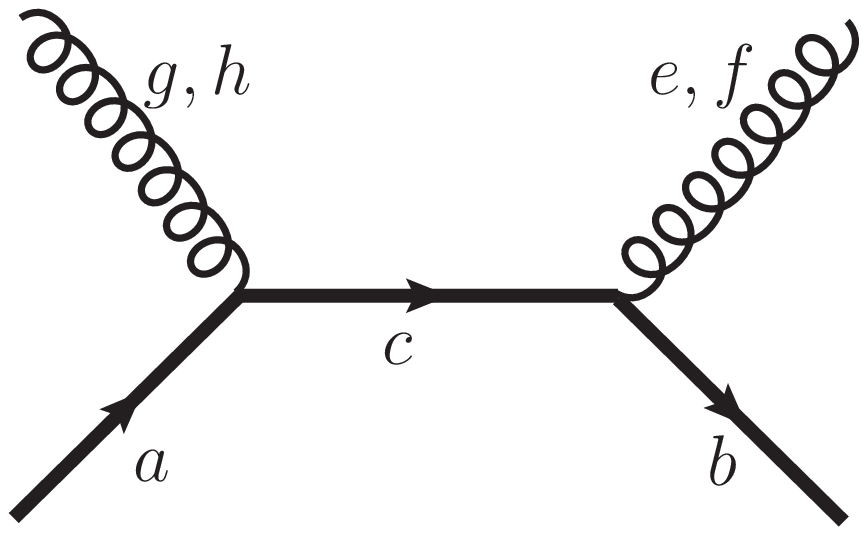}} 
\caption{s-channel Compton scattering (left). u-channel Compton scattering (right)}
\label{su}
\end{figure}

\textbf{Compton scattering:} There are three types of scatterings ($s,t$ and $u$ channels) that contribute to the Compton scattering i.e., scattering off of a gluon from a quark. For $s$ and $u$ channel scatterings, the corresponding Feynmann diagrams are shown in Fig.[\ref{su}] and for the $t$ channel scattering the relevant diagram is shown in right side of Fig.[\ref{st}]. We shall evaluate the scattering amplitude for Compton scattering below.

\textbf{s-channel:} The relevant diagram for this channel is shown on the left side of Fig.[\ref{su}] where $ef(gh), a(b)$ are color indices of incoming (outgoing) gluon and quark. In the double line notation, the scattering amplitude for the process is given as
\begin{equation}
i\mathcal{M}_{s}=i{(ig)^2}(t^{ef})_{ac}(t^{gh})_{cb}\bigg[\frac{\bar{u}_{b}(p')\slashed{\epsilon}(\slashed{p}+\slashed{q}+M)\slashed{\epsilon}u_{a}(p)}{s-M^2}\bigg].
\label{ms}
\end{equation}
where $s$ is Mandelstam variable and $M$ is the mass of HQ. Note here that unlike Coulomb scattering there is no color dependence on the HQ propagator. This is because of the large mass of heavy quark. For massive quark and massless gluon the matrix element squared for s-channel Compton scattering can be written as 
\begin{equation}
|\mathcal{M}_s|^{2}=\frac{8g^4}{16 N_c(N_c^2-1)}\mathcal{P}^{ef}_{ac}\mathcal{P}^{ef}_{ac'}\mathcal{P}^{gh}_{cb}\mathcal{P}^{gh}_{c'b}\bigg(\frac{M^2(M^2-u-3s)-us}{(s-M^2)^2}\bigg).
\label{ms1}
\end{equation}
Note that the scattering amplitude  depends on the color of quarks and gluons. For the evaluation of the transport coefficients one needs to perform a color sum. Same as in the case of Coulomb scattering, the color indices of incoming and outgoing gluon ($ef,gh$) in Eq.(\ref{ms1})  will be summed with the distribution functions appearing in Eq.(\ref{integral1}).

{\textbf{u channel:}} The corresponding Feynman diagram for $u$ channel Compton scattering is illustrated at the right side of Fig.(\ref{su}). Scattering amplitude that depends on the color of incoming and outgoing color particles can be written as
\begin{equation}
i\mathcal{M}_{u}=i(ig)^2(t^{ef})_{cb}(t^{gh})_{ac}\bigg[\frac{\bar{u}_{b}(p')\slashed{\epsilon}(\slashed{p}-\slashed{q'}+M)\slashed{\epsilon}u_{a}(p)}{u-M^2}\bigg].
\label{uch}
\end{equation}
Simplifying Eq.(\ref{uch}) with the polarization sum of massless gluon and spin sum and average over heavy quark gives
\begin{equation}
|\mathcal{M}_u|^{2}=\frac{8g^4}{16N_c(N_c^2-1)}\mathcal{P}^{gh}_{ac}\mathcal{P}^{gh}_{ac'}\mathcal{P}^{ef}_{cb}\mathcal{P}^{ef}_{c'b}\bigg(\frac{M^2(M^2-3u-s)-us}{(u-M^2)^2}\bigg).
\label{uch1}
\end{equation}
In the Eq.(\ref{uch1}), the product of projection operator can be simplified by summing over the color indices $a,b$  and $c$. However, the color indices of initial and final gluon should be summed with the distibution function in Eq.(\ref{integral1}). Keeping $ef,gh$ as open indices, the product of the projection operators  can be simplified to
\begin{eqnarray}
\mathcal{P}^{gh}_{ac}\mathcal{P}^{gh}_{ac'}\mathcal{P}^{ef}_{cb}\mathcal{P}^{ef}_{c'b}&=&\delta_{eh}-\frac{1}{N_c}\bigg(2\delta_{ef}\delta_{fh}\delta_{eh}+\delta_{gh}\delta_{eg}\delta_{eh}\bigg)+\frac{1}{N_c^2}\bigg(\delta_{ef}+\delta_{ef}\delta_{gf}\delta_{eh}\delta_{gh}+\delta_{ef}\delta_{fh}\delta_{eg}\delta_{gh}\nonumber\\
&+&\delta_{eh}\delta_{fg}\delta_{ef}\delta_{gh}+\delta_{gh}\bigg)-\frac{1}{N_c^3}\bigg(\delta_{ef}\delta_{gh}+\delta_{ef}\delta_{eh}\delta_{gh}\bigg).
\end{eqnarray}

{\textbf{t channel:}} The relevant Feynmann diagram for the $t$ channel Compton scattering is shown on the right side of Fig.[\ref{st}]. For the color dependent scattering amplitude one can write
\begin{equation}
i\mathcal{M}_{t}=(ig)^2(t^{ml})_{ab}f^{cd,ef,gh}\bigg[\frac{\epsilon_{\mu}(q)\epsilon^{*}_{\nu}(q')C^{\mu \alpha \nu}(q-q',-q,-q')\bar{u}_{b}(p')\gamma_{\alpha}u_{a}(p)}{(t+(m_D^2)_{mlcd})}\bigg],
\label{compt}
\end{equation}
where 
\begin{equation}
C^{\mu \nu \rho}(k_1,k_2,k_3)=[(k_1-k_2)^{\rho}g^{\mu \nu}+(k_2-k_3)^{\mu}g^{\nu \rho}+(k_3-k_1)^{\nu}g^{\mu \rho}].
\label{cmu}
\end{equation}
In Eq.(\ref{compt}), $f^{cd,ef,gh}$ is structure constant as defined in Eq.(\ref{structure}) and $\epsilon_{\mu}(q),\epsilon^{*}_{\nu}(q')$ are the polarization vectors for incoming and outgoing gluon. The matrix element squared can be obtained by performing appropriate polarization sum for gluons and spin sum and average for heavy quark. Doing so, matrix element squared becomes
\begin{equation}
|\mathcal{M}_t|^{2}=\frac{16g^4}{8N(N^2-1)}\mathcal{P}^{ml}_{ab}\mathcal{P}^{l'm'}_{ba}f^{cd,ef,gh}f^{d'c',fe,hg}\bigg(\frac{-(M^2-s)(M^2-u)}{(t+(m_D^2)_{mlcd})t+(m_D^2)_{m'l'c'd'}}\bigg).
\end{equation}
The corresponding interference terms among Compton scatterings are given in appendix(A). To that the total scattering amplitude of Compton scattering that enters in Eq.(\ref{integral1}) for evaluation of the drag and the diffusion coefficients is $|\mathcal{M}_{gQ}|_{efgh}^2=|\mathcal{M}_s|^2+ |\mathcal{M}_u|^2+|\mathcal{M}_t|^2+|\mathcal{M}_s|^{\dagger}|\mathcal{M}_u|+|\mathcal{M}_{u}|^{\dagger}|\mathcal{M}|_s+|\mathcal{M}_s|^{\dagger}|\mathcal{M}_t|+|\mathcal{M}_{t}|^{\dagger}|\mathcal{M}|_s|+|\mathcal{M}_t|^{\dagger}|\mathcal{M}_u|+|\mathcal{M}_{u}|^{\dagger}|\mathcal{M}|_t$.  For computational simplification, we shall use the leading order contribution in the Debye mass that appears in the $t$ channel scatterings. 
\section{Results and discussions}
\label{results}
With the scattering amplitude for the processes $l Q \rightarrow l Q$ (where $l$ stands for light quark/antiquark and gluon and $Q$ stands for HQ) as evaluated in the previous section, we numerically compute the drag and the diffusion coefficients using Eq.[\ref{integral1}] and incorporate the dissipative effects in the quark/antiquark and gluon color distribution functions as defined in Eqs.(\ref{quarkdis}) and (\ref{gluondis}). 
\begin{figure}[tbh]
\subfigure{
\includegraphics[width=6.9cm]{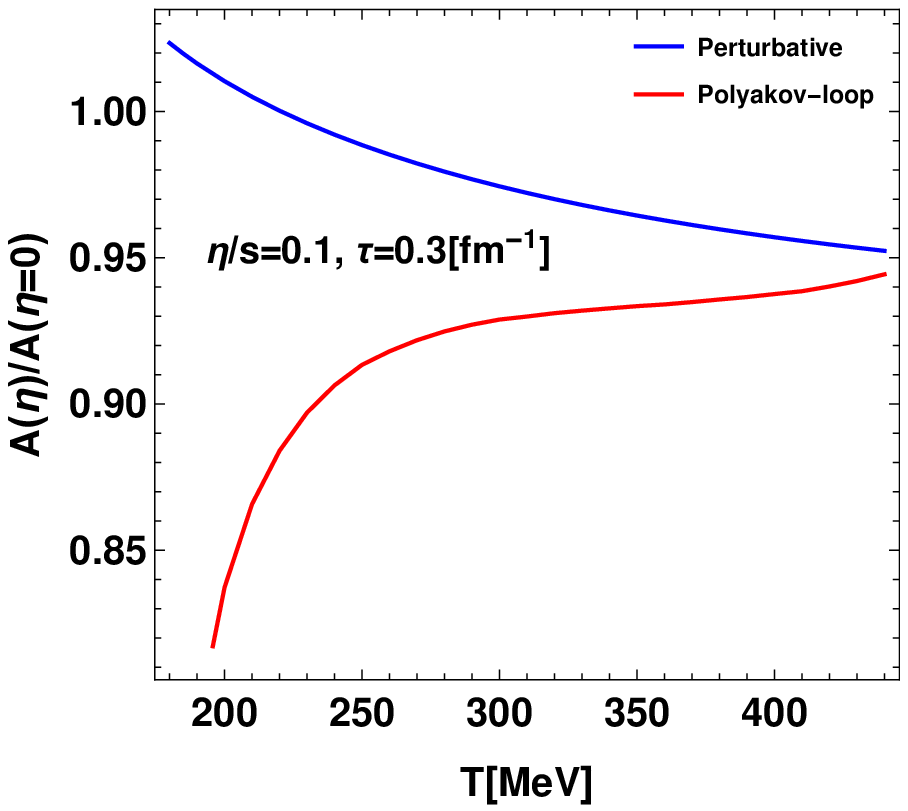}}
\subfigure{
\hspace{-0mm}\includegraphics[width=6.8cm]{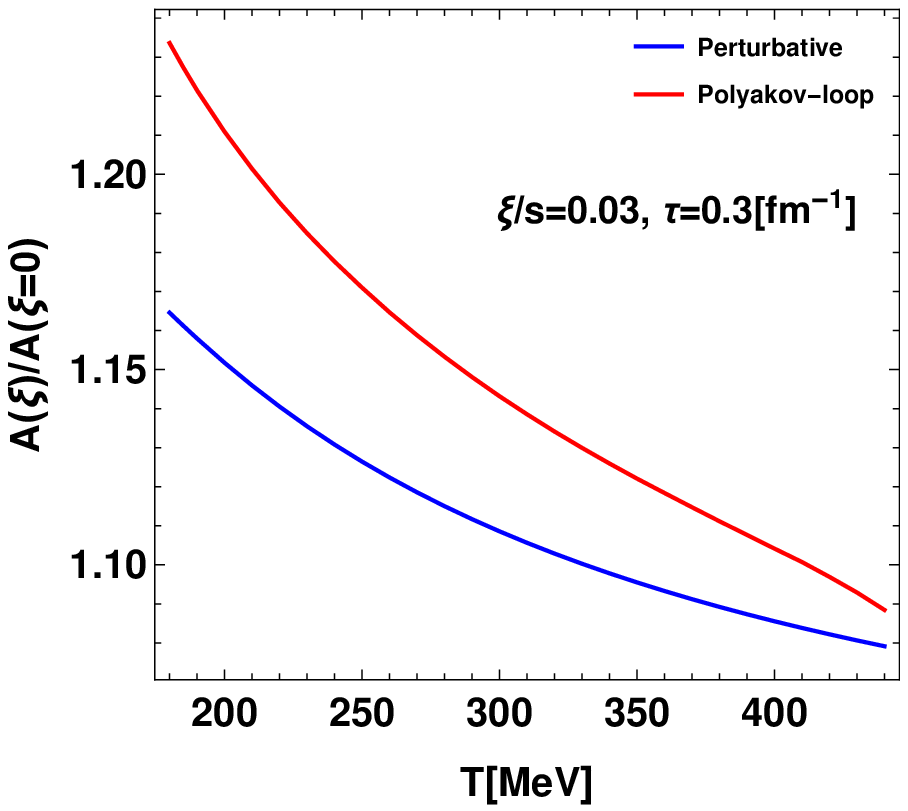}} 
\caption{\textbf{Left panel:} The ratio of the drag coeffient i.e., $A(\eta)/A(\eta=0)$ is shown as a function of temperature for $p=1$ GeV, $\eta/s=0.1$, $\tau=0.3$ fm$^{-1}$ and $\xi/s=0$. The blue curve represents the perturbative QCD result while the red curve correspond to including Polyakov loop withing the matrix model. \textbf{Right panel:} The ratio $A(\xi)/A(\xi=0)$ is shown as a function of temperature for $\xi/s=0.03$, $p=1$ GeV, $\tau=0.3$ fm$^{-1}$ and $\eta/s=0$. Similar to the left panel, the blue curve correspond to pQCD results and the red curve correspond to including Polyakov loop within the matrix model.}
\label{dragT}
\end{figure}
For this purpose, we use charm quark mass $M=1.27$ GeV and the two loop running coupling constant~\cite{Caswell:1974gg}
\begin{equation}
\alpha_s=\frac{1}{4 \pi} \frac{1}{2 \beta_0 \ln\frac{\pi T}{\Lambda}+\frac{\beta_1}{\beta_0}\ln(2\ln(\frac{\pi T}{\Lambda}))}
\end{equation} 
where 
\begin{equation}
\beta_{0}=\frac{1}{16 \pi^2}\bigg(11-\frac{2 N_f}{3}\bigg)
\end{equation}
\begin{equation}
\beta_{1}=\frac{1}{(16 \pi^2)^2}\bigg(102-\frac{38 N_f}{3}\bigg)
\end{equation}
with $\Lambda=260$ MeV and light quark flavor $N_f=2$. We also evaluate the HQ transport coefficients in pQCD by evaluating scattering amplitude squared withing the pQCD framework.
\begin{figure}[tbh]
\subfigure{
\includegraphics[width=6.8cm]{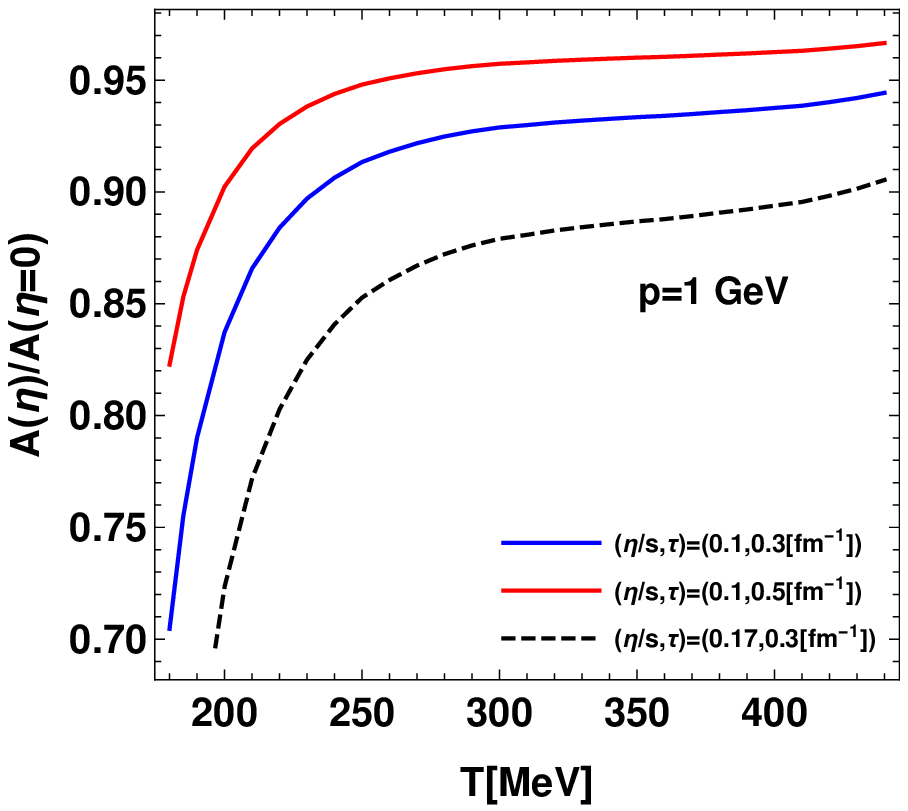}}
\subfigure{
\hspace{-0mm}\includegraphics[width=6.8cm]{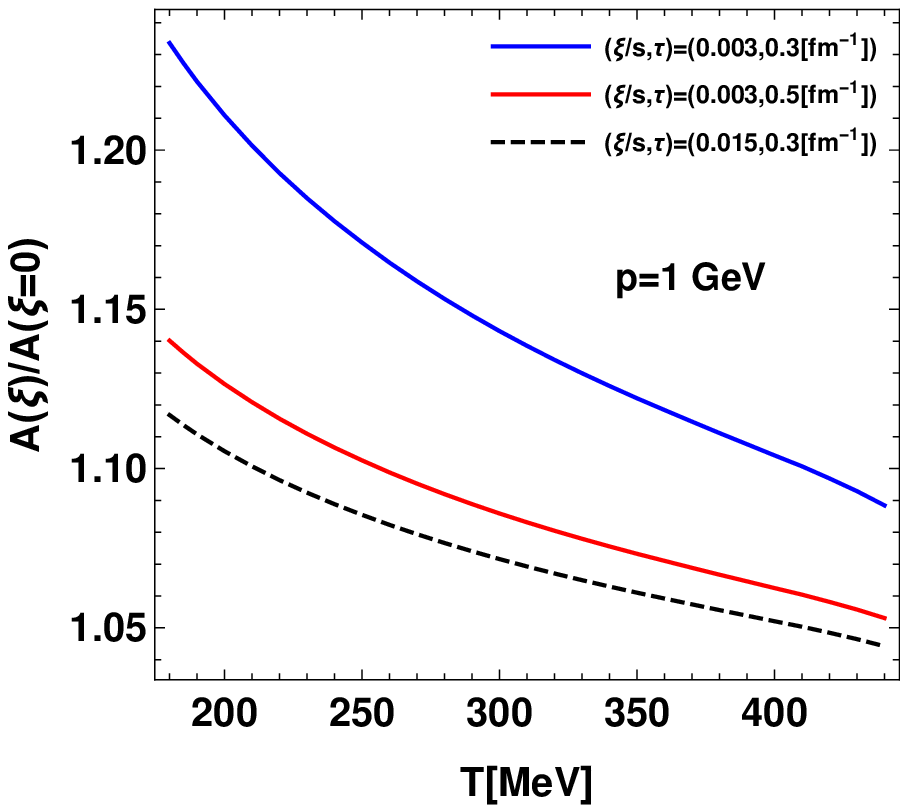}} 
\caption{\textbf{Left panel:} The ratio $A(\eta)/A(\eta=0)$ as a function of temperature for HQ momentum $p=1$ GeV and $\xi/s=0$. The topmost curve is for $\eta/s=0.1, \tau=0.3$ fm$^{-1}$, bottom-most i.e., dashed black cuve is for $\eta/s=0.17, \tau=0.3$ fm$^{-1}$ and the blue curve is for $\eta/s=0.1, \tau=0.5$ fm$^{-1}$. \textbf{Right panel:} The ratio of the drag coefficient $A(\xi)/A(\xi=0)$ as a function of temperature for HQ momentum $p=1$ GeV and $\eta/s=0$. The blue curve  correspond to $\xi/s=0.03,\tau=0.3$ fm$^{-1}$, the red curve correspond to $\xi/s=0.03,\tau=0.5$ fm$^{-1}$ and the dashed black curve correspond to $\xi/s=0.015,\tau=0.3$ fm$^{-1}$.}
\label{dragT1}
\end{figure}

\begin{figure}[tbh]
\subfigure{
\includegraphics[width=6.8cm]{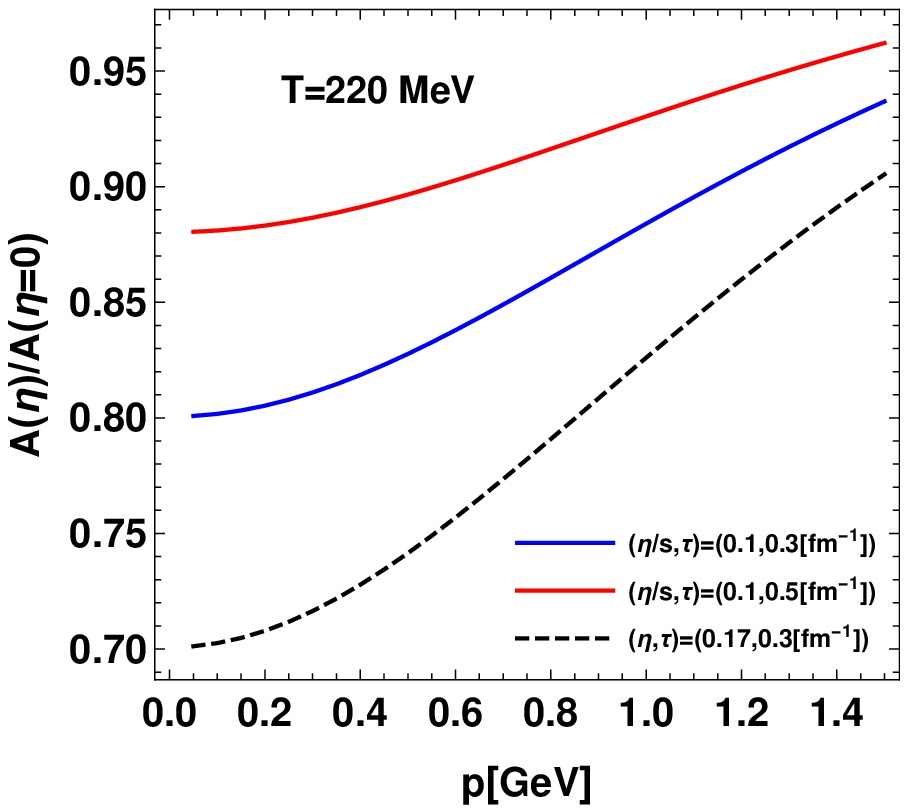}}
\subfigure{
\hspace{-0mm}\includegraphics[width=6.8cm]{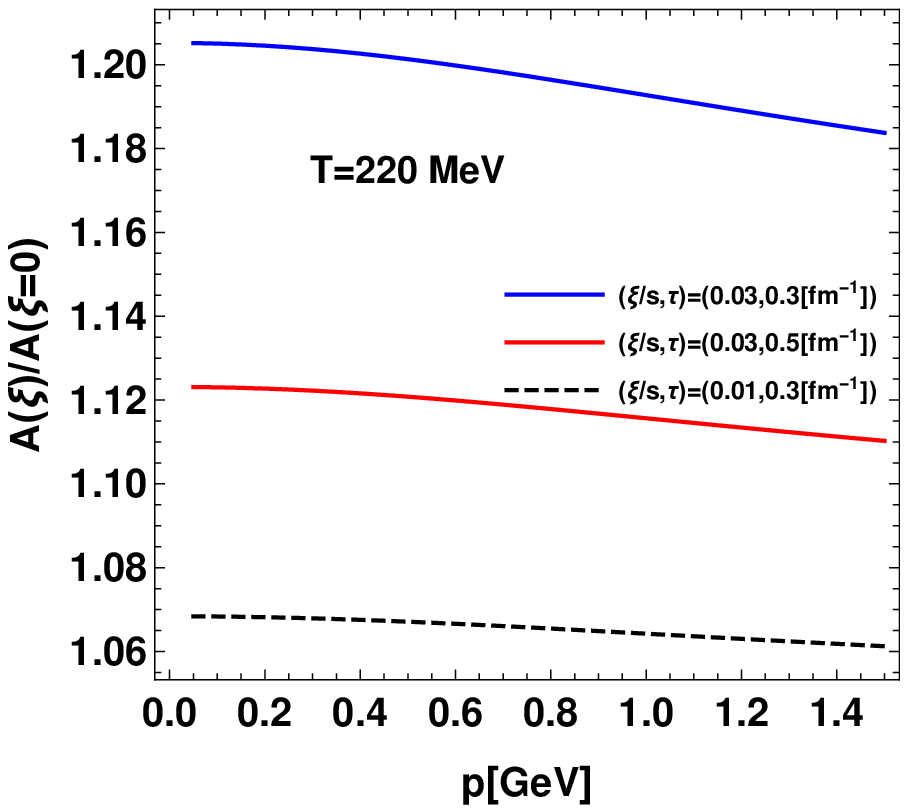}} 
\caption{\textbf{Left panel:} Variation of $A(\eta)/A(\eta=0)$ as a function of momentum for $\xi/s=0$ and $T=220$ MeV. The red curve correspond to $\eta/s=0.1,\tau=0.5$ fm$^{-1}$, the blue curve correspond to $\eta/s=0.1,\tau=0.3$ fm$^{-1}$ and the dashed black curve correspond to $\eta/s=0.17,\tau=0.3$ fm$^{-1}$. \textbf{Right panel:} Variation of $A(\xi)/A(\xi=0)$ as a function of momentum for $\eta/s=0$ and $T=220$ MeV. The blue curve correspond to $\xi/s=0.03,\tau=0.3$ fm$^{-1}$, the red curve correspond to $\xi/s=0.03,\tau=0.5$ fm$^{-1}$ and the dashed black curve correspond to $\xi/s=0.01,\tau=0.3$ fm$^{-1}$.}
\label{dragp}
\end{figure}

In general, there are two factors that essentially affect the heavy quark transport properties. One is the Debye mass that occurs in the evaluation of matrix elements and the other is the Polyakov loop dependent distribution functions of quark/anti-quark and gluon. At low temperature, a lower value of the Debye mass increases the transport coefficients. On the other hand, the distribution function with the non-trivial $\phi$ tend to reduce it. These apart a third factor that plays an important role here is the momentum dependence of departure $\delta f_{a/ab}$ in Eqs.(\ref{quarkdis}) and (\ref{gluondis}) of the distribution function from the equiliberium distribution function.  Now let us examine the results in somewhat detail. In Fig.(\ref{dragT}), we show the dependence of the drag coefficient as a function of temperature. In the left panel, we have plotted the drag coefficient i.e., Eq.(\ref{drag}) for a constant value of $\eta/s$ normalized to the drag coefficient for $\eta/s=0,\xi/s=0$. In both the figures, we have taken a value of $\tau=0.3$ fm$^{-1}$ and $p$, the magnitude of the momentum of HQ as $p=1$ GeV. To discuss the effects of shear viscosity ($\eta$), we have plotted $A(\eta)/A(\eta=0)$ as a function of temperature. The blue curve corresponds to the pQCD calculation results and the red curve corresponds to the effect of the Polyakov loop within the matrix model. It is clear that at low temperature for $\eta/s=0.1$ and $\tau=0.3$ fm$^{-1}$, the drag coefficient is smaller with the Polyakov loop as compared to pQDC. This is mainly because of the negative contribution from the momentum factor ($q^2/3-q_z^2$) in $\delta f_{a/ab}$ along with the effect of Polyakov loop. Such a non-monotonic behavior with the  Polyakov loop can be understood as follows. The Debye mass is smaller at lower temperature due to the small value of Polyakov loop compared to the high temperatur~\cite{Singh:2018wps}. As the temperature increases the suppression decreases and approach the perturbative results beyond which it decreases similar to the perturbative QCD results. Another reason for more suppression in the drag within the matrix model is due to the distribution function i.e., colored particles are suppressed due to small value of Polyakov loop compared to pQCD. In the right panel Fig.(\ref{dragT}),  the temperature behaviour of the drag with the inclusion of bulk viscosity is shown for constant $\xi/s$ and $\eta/s=0$ .  The drag is more with the inclusion of the bulk viscosity compared to the $\xi/s=0$ case. This is because $\xi/s$ term in $\delta f_{a/ab}$ is always positive and there is no phase space suppression so the small Debye mass enhances the drag coefficient. In the matrix model drag is more as compared to pQCD due to small value Debye mass and distribution functions. 

The ratio $A(\eta)/A(\eta=0)$ of the drag coefficient as defined in Eq.(\ref{drag}) is plotted as a function of temperature for HQ momentum $p=1$ GeV in Fig.(\ref{dragT1}) for various value of $\eta/s$ and $\tau$ to see the effect of both ($\eta/s,\tau$) the quantities. Here the scattering amplitude squared for the relevant scatterings are evaluated within the matrix model. As anticipated from the effect of phase space, Polyakov loop dependent distribution functions of quark/anti-quark and gluon and the Debye mass, with an increase in $\eta/s$ value, the HQ drag coefficient decreases as shown on the left side of Fig.(\ref{dragT1}). Here the red line is for $\eta/s=0.1$ and black dashed line for $\eta/s=0.17$ with $\tau=0.3$ fm$^{-1}$. For small value of $\eta/s$ and sufficiently large value of thermalization time ($\tau$), the effect of $\eta/s$ on the HQ drag is weak as shown by the red curve. This can be understood from $1/\tau$ factor in Eqs.[\ref{qrkdist}] and [\ref{gludist}]. On the right side of Fig.(\ref{dragT1}), the effect of $\xi/s$ and $\tau$ on the drag coefficient is shown. As expected, with an increase in $\xi/s$, the drag coefficient increases as shown by a dashed black line i.e., $\xi/s=0.015$ and the blue line with $\xi/s=0.03$. Same as earlier, with an increase in thermalization time, the drag coefficient decreases.

\begin{figure}[tbh]
\subfigure{
\includegraphics[width=6.8cm]{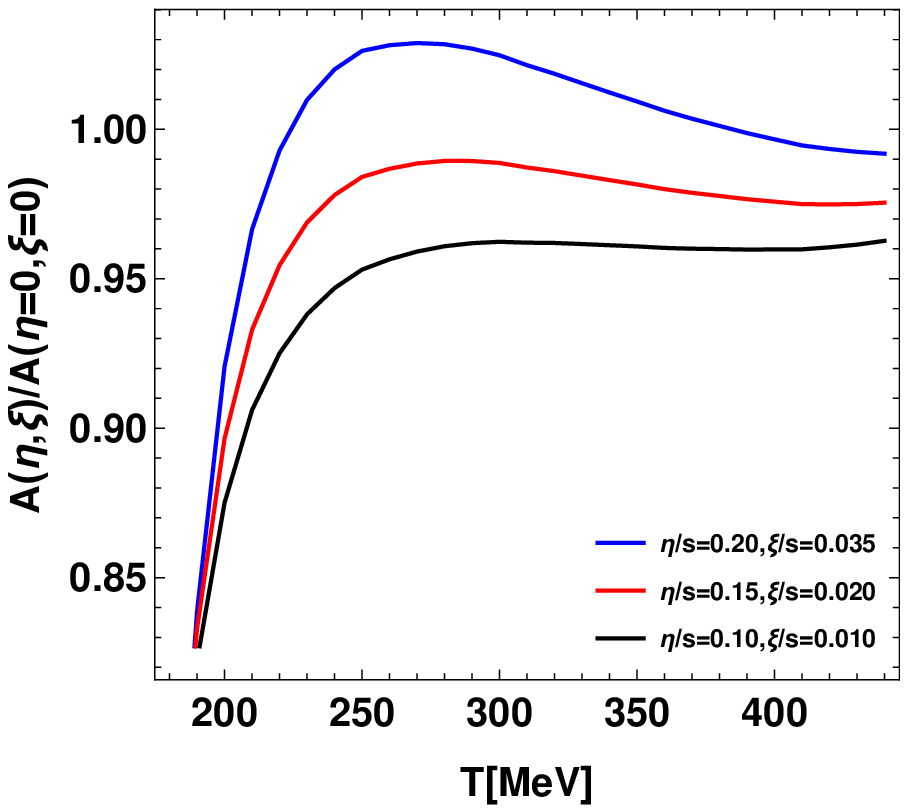}}
\subfigure{
\hspace{-0mm}\includegraphics[width=6.8cm]{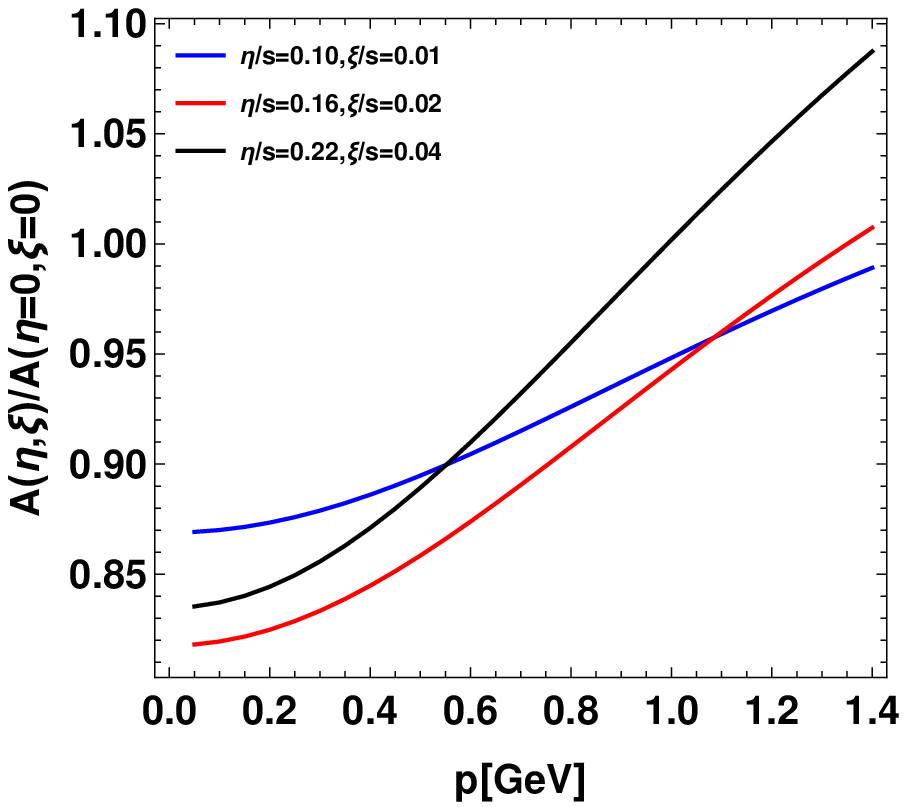}} 
\caption{\textbf{Left panel:} The ratio $A(\eta,\xi)/A(\eta=0,\xi=0)$ as a function of temperature for HQ momentum $p=1$ GeV and $\tau=0.3$ fm$^{-1}$. The blue curve correspond to $\eta/s=0.2, \xi/s=0.035$, the red curve correspond to $\eta/s=0.15, \xi/s=0.020$ and the black curve correspond to $\eta/s=0.1, \xi/s=0.01$. \textbf{Right panel:} The ratio $A(\eta,\xi)/A(\eta=0,\xi=0)$ as a function of heavy quark momentum for $T=220$ MeV and $\tau=0.3$ fm$^{-1}$. The blue curve correspond to $\eta/s=0.1, \xi/s=0.01$, the red curve correspond to $\eta/s=0.15, \xi/s=0.020$ and the black curve correspond to $\eta/s=0.22, \xi/s=0.04$.}
\label{dragp1}
\end{figure}

\begin{figure}[tbh]
\subfigure{
\includegraphics[width=6.8cm]{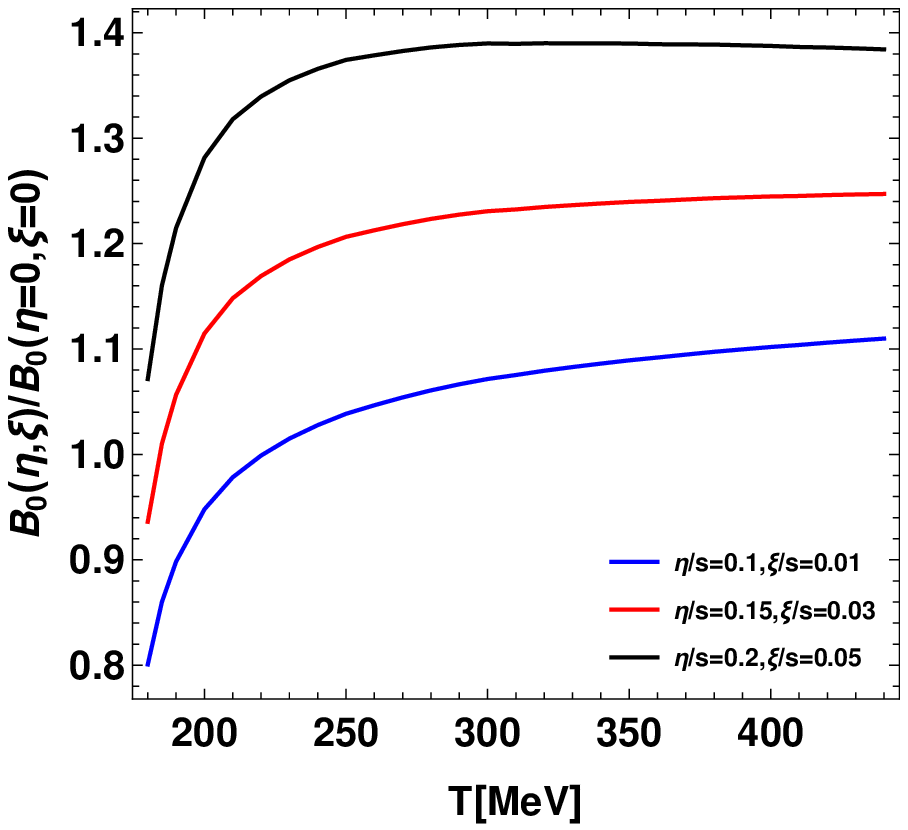}}
\subfigure{
\hspace{-0mm}\includegraphics[width=6.8cm]{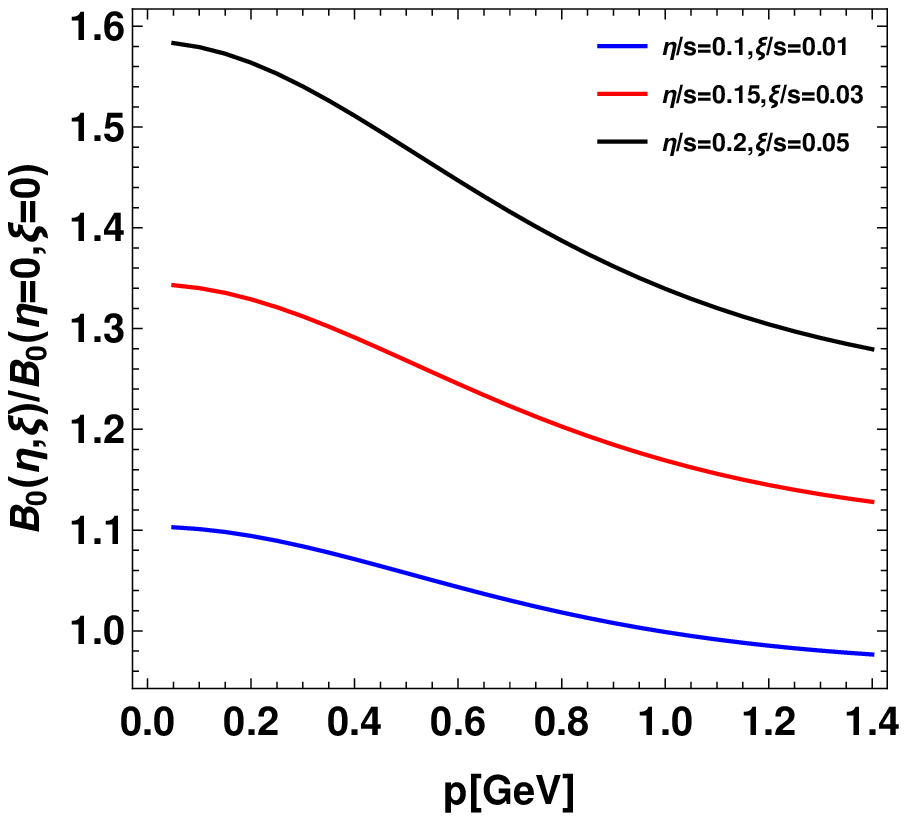}} 
\caption{\textbf{Left panel:} The ratio $B_{0}(\eta,\xi)/B_{0}(\eta=0,\xi=0)$ of diffusion coefficient  as a function of temperature for HQ momentum $p=1$ GeV and $\tau=0.3$ fm$^{-1}$. The black curve correspond to $\eta/s=0.1, \xi/s=0.01$, the red curve correspond to $\eta/s=0.15, \xi/s=0.03$ and the blue curve correspond to $\eta/s=0.2, \xi/s=0.05$. \textbf{Right panel:} The ratio $B_{0}(\eta,\xi)/B_{0}(\eta=0,\xi=0)$ of diffusion coefficient  as a function of HQ momentum for temperature $T=220$ MeV and $\tau=0.3$ fm$^{-1}$. The black curve correspond to $\eta/s=0.1, \xi/s=0.01$, the red curve correspond to $\eta/s=0.15, \xi/s=0.03$ and the blue curve correspond to $\eta/s=0.2, \xi/s=0.05$.}
\label{diff}
\end{figure}

Drag coefficient $A(\eta)$ normalized with $A(\eta=0)$ as a function of HQ momentum $p$ for $T=220$ MeV is shown in Fig.(\ref{dragp}). On the left side of the same figure, for finite shear viscosity to entropy ratio, with an increase in momentum, the drag coefficient increases. However with increase in this $\eta/s$ the drag coefficient decreases as shown by the black dashed line. As earlier, this behaviour can be interpreted by the phase space suppression and the Debye mass. Similarly, with an increase in thermalization time, the drag coefficient decreases as can be anticipated from Eqs.(\ref{quarkdis}) and (\ref{gluondis}).  On the right side of the same figure the effect of bulk viscosity on HQ drag is shown. Here, the drag coefficient decreases with an increases in HQ momentum. Note that unlike $\eta/s$, with increase in $\xi/s$ the drag coefficient increases, however with an increases in thermalization time the drag coefficient decreases. 

On the left side of Fig.(\ref{dragp1}), the effect of both $\xi/s$ and $\eta/s$ for $\tau=0.3$ fm$^{-1}$ on the ratio $A(\eta,\xi)/A(\eta=0,\xi=0)$ of the drag coefficient as a function of temperature is shown. At low temperature, shear viscosity dominates due to the phase space suppression, so the drag coefficient decreases.  At moderate temperature e.g., around 250 MeV bulk viscosity dominates so the drag increases. Again at high temperature around 320 MeV, both the $\eta/s$ and $\xi/s$ decreases the drag coefficient. For small values of $\eta/s$ and $\xi/s$ i.e.,  $\eta/s=0.1, \xi/s=0.01$, the dependence of the drag coefficient on the medium temperature is somewhat weak, however, it grows for a larger values of $\eta/s$ and $\xi/s$. On the right side of Fig.(\ref{dragp1}) the same ratio as a function of momentum is shown. Similar to the case of temperature behavior, for smaller values of $\eta/s$ and $\xi/s$ the drag coefficient is somewhat weakly dependent on momentum (see blue curve; $\eta/s=0.1, \xi/s=0.01$), however, it strongly depends on momentum for larger values $\eta/s$ and $\xi/s$.. At low momentum, the drag coefficient suppresses, however, on the other hand at high momentum drag coefficient enhances for any constant values of $\eta/s$ and $\xi/s$. 

The ratio $B_{0}(\eta,\xi)/B_{0}(\eta=0,\xi=0)$ of diffusion coefficients as defined in Eq.(\ref{diffusion}) is plotted as a function of temperature and momentum  in Fig.(\ref{diff}). On the left side of Fig.(\ref{diff}), the black curve is for $\eta/s=0.2, \xi/s=0.05$, the red curve is for $\eta/s=0.15, \xi/s=0.03$ and the blue curve correspond to $\eta/s=0.1, \xi/s=0.01$ for HQ momentum $p=1$ GeV. Similar to the drag coefficient, at lower values of  $\eta/s$ and $\xi/s$ e.g., $\eta/s=0.1, \xi/s=0.01$ the diffusion coefficient is not affected much. However, at larger values e.g.,$\eta/s=0.2, \xi/s=0.05$ the diffusion coefficient enhances. At low temperature, with the inclusion of dissipative effects, the diffusion coefficient is small compared to the case of without vicous effetcs, however, at high temperature the trend is quite opposite. On the right hand side of the same figure, momentum dependence has opposite trend as compared to that of  temperature. At low momentum, with the inclusion of disspative effects, the diffusion coefficient enhances as compared to the case of without viscous effects and at high momentum dissipative effects reduces the diffusion coefficient. In pQCD calculation, the Drag and the diffusion coefficients for various values of $\eta/s, \xi/s, \tau$ as a function of temperature and momentum that are presented here behave similar to as pointed out in Ref.\cite{Das:2012ck}. However, the differences are due to the effect of the Polyakov loop.   
\section{Summary} 
\label{summary}
In the present work, we have computed the corrections due to the effects of the shear and the bulk viscosities on the HQ drag and diffusion coefficients within the matrix model of semi QGP. To incorporate the viscous corrections we first write the distribution function of quark and gluon ($f_{a/cd}=f^{0}_{a/cd}+\delta f_{a/cd}$, where $f^{0}_{a/cd}$ is equilibrium distribution function and $\delta f_{a/cd}$ summarises the effect of shear and bulk viscosities)  as defined in Eqs.(\ref{quarkdis}) and (\ref{gluondis}) using second moment ansatz. We next calculate the color dependent scattering amplitude of HQ from the light thermal partons in the bulk medium within the matrix model of semi QGP. Non-perturbative effects are included via the Polyakov loop in quark/antiquark and gluon distribution functions as well as in the Debye mass. In all the calculations, we have taken the constant values for the viscosity to entropy density ratio .e., without their temperature dependence. However, these can be calculated as has been done in Ref.{\cite{Hidaka:2009ma}}.  With a reasonable constant value for $\eta/s$ and $\xi/s$ for the temperature range we have considered, we find that the drag coefficient withing the matrix model is small compared to that of perturbative QCD. Similarly, for a constant value of $\xi/s$, the drag coefficients is large withing the matrix model compared the pQCD resutls. Furthermore, with an increase in temperature and momentum the drag coefficient increases, however, the diffusion coefficient increases with an increase in temperature and decreases with an increase in momentum.  In addition, for small $\eta/s$ and $\xi/s$, both the drag and the diffusion coefficients have a weak dependence on temperature and momentum for all range of temperature and momentum considered here.   

\appendix
\section{Interference terms in the scattering amplitude}
\label{appendixA}
In this section, we shall discuss the interference amplitudes of $s$, $t$ and $u$ channel scatterings contributing to Compton scattering. For $s$ and $u$ channel of the scatterings,  the interference term can be written as
\begin{equation}
\mathcal{M}_{s}{\mathcal{M}_u}^\dagger=g^4 (t^{ef})_{ac}(t^{gh})_{cb}(t^{gh})_{ac'}(t^{ef})_{c'b}\bigg[\frac{Tr[(\slashed{p'}+M)\gamma^{\nu}(\slashed{p}+\slashed{q}+M)\gamma^{\mu}(\slashed{p}+M)\gamma^{\alpha}(\slashed{p}-\slashed{q'}+M)\gamma^{\beta}]\epsilon_{\mu}(q)\epsilon_{\beta}(q)\epsilon^{*}_{\nu}(q')\epsilon^{*}_{\alpha}(q')}{(s-M^2)(u-M^2)}\bigg]
\label{suit}
\end{equation}
where the trace is over Dirac matrices and the terms like $(t^{ab})_{cd}$ are generators of the group which can be written in terms of projection operators as given in Eq.(\ref{pop}). Using the polarization sum for massless gluons, and spin sum/average of final/initial quark, one can simplify Eq.(\ref{suit}) to yield
\begin{equation}
\mathcal{M}_{s}{\mathcal{M}_u}^\dagger=\frac{g^4}{16N_c(N_c^2-1)}\mathcal{P}^{ef}_{ac}\mathcal{P}^{gh}_{cb}\mathcal{P}^{gh}_{ac'}\mathcal{P}^{ef}_{c'b} \bigg(\frac{-8M^2(t-4M^2)}{(s-M^2)(u-M^2)}\bigg).
\label{suit1}
\end{equation}
As earlier,  color index $ef/gh$ of incoming/outgoing gluon will be summed with the statistical distribution function while evaluating the drag and the diffusion coefficients using Eq.(\ref{integral1}). Similarly, another term that contributes to the amplitude i.e., $\mathcal{M}_{s}^\dagger{\mathcal{M}_u}$ of the interference of the same scattering channels is given as
\begin{equation}
\mathcal{M}_{s}^\dagger{\mathcal{M}_u}=\frac{g^4}{16N_c(N_c^2-1)}\mathcal{P}^{ef}_{ac}\mathcal{P}^{gh}_{cb}\mathcal{P}^{gh}_{ac'}\mathcal{P}^{ef}_{c'b} \bigg(\frac{-8M^2(t-4M^2)}{(s-M^2)(u-M^2)}\bigg).
\end{equation}
For $s$ and $t$ channel scatterings, the matrix element squared of the interference term can be given as
\begin{equation}
\mathcal{M}_{s}{\mathcal{M}_t}^\dagger=g^4 (t^{ef})_{ac}(t^{gh})_{cb}(t^{lm})_{ba}f^{dc,fe,hg}\bigg[\frac{Tr[(\slashed{p'}+M)\gamma^{\nu}(\slashed{p}+\slashed{q}+M)\gamma^{\mu}(\slashed{p}+M)\gamma^{\lambda}C^{\alpha \beta \sigma}(q-q',-p,p')g_{\alpha \beta}]\epsilon_{\mu}\epsilon_{\alpha}\epsilon^{*}_{\nu}\epsilon^{*}_{\sigma}}{(s-M^2)(t+(m_D^2)_{mlcd})}\bigg],
\label{stit}
\end{equation}
where $C^{\mu \nu \sigma}$ is defined in Eq.(\ref{cmu}). With the polarization sum of massless gluon and trace over Dirac space, the scattering amplitude can be simplified to
\begin{equation}
\mathcal{M}_{s}{\mathcal{M}_t}^\dagger=\mathcal{M}_{s}^\dagger{\mathcal{M}_t}=\frac{g^4\sqrt{2}}{16N_c(N_c^2-1)}\mathcal{P}^{ef}_{ac}\mathcal{P}^{gh}_{cb}\mathcal{P}^{lm}_{ba}f^{dc,fe,hg} \bigg(\frac{16M^2s-8M^4-us}{(s-M^2){(t+(m_D^2)_{mlcd})}}\bigg).
\end{equation}
The last term contributing to the scattering amplitude of the intefering diagrams comes from the $u$ and $s$ channel scatterings and can be given as
\begin{equation}
\mathcal{M}_{u}{\mathcal{M}_t}^\dagger=g^4 (t^{gh})_{ac}(t^{ef})_{cb}(t^{lm})_{ba}f^{dc,fe,hg}\bigg[\frac{Tr[(\slashed{p}+M)\gamma^{\mu}(\slashed{p'}-\slashed{q}+M)\gamma^{\nu}(\slashed{p}+M)\gamma^{\lambda}C^{\alpha \beta \sigma}(q-q',-q,-q')g_{\lambda \alpha}]\epsilon_{\mu}\epsilon_{\alpha}\epsilon^{*}_{\nu}\epsilon^{*}_{\sigma}}{(u-M^2)(t+(m_D^2)_{mlcd})}\bigg].
\label{utit}
\end{equation}
Similar to Eqs.(\ref{suit}) and (\ref{stit}), Eq.(\ref{utit}) can be simplified to 
\begin{equation}
\mathcal{M}_{u}{\mathcal{M}_t}^\dagger=\mathcal{M}_{u}^\dagger{\mathcal{M}_t}=\frac{g^4\sqrt{2}}{16N_c(N_c^2-1)}\mathcal{P}^{gh}_{bc}\mathcal{P}^{ef}_{ca}\mathcal{P}^{lm}_{ab}f^{dc,fe,hg} \bigg(\frac{32M^4-8M^2t}{(u-M^2){((t+(m_D^2)_{mlcd}))}}\bigg).
\end{equation}


\end{document}